\documentclass[aps,10pt,reprint,showpacs,superscriptaddress,longbibliography]{revtex4-2}
\pdfoutput=1 
\usepackage[usenames,dvipsnames]{xcolor}
\usepackage[bookmarks=true,colorlinks,citecolor=blue,urlcolor=blue]{hyperref}
\usepackage{amssymb}
\usepackage{graphicx}
\usepackage{epsfig}
\usepackage{epstopdf}
\usepackage{mathtools}
\usepackage{braket}
\usepackage{blindtext}
\usepackage[compat=1.1.0]{tikz-feynman}
\usepackage[normalem]{ulem}
\usepackage[english, russian]{babel}
\usepackage[toc,page,header]{appendix}
\usepackage{minitoc}
\usepackage{tocloft}
\usepackage[cp1251]{inputenc}

\definecolor{pink2}{rgb}{0.858, 0.188, 0.478}

\begin{document}

\doparttoc 
\faketableofcontents 


\title{On the Possibility of a Significant Increase in the Storage Time
	of Ultracold Neutrons in Traps Coated with a Liquid Helium Film}

\newcommand{\LND}{\affiliation{Landau
		Institute for Theoretical Physics, Russian Academy of Sciences, Chernogolovka, Moscow region, 142432 Russia}}
\newcommand{\MSS}{\affiliation{National University of Science and Technology MISiS, Moscow, 119049 Russia}}
\newcommand{\NRCK}{\affiliation{National Research Center Kurchatov Institute, Moscow, 123182 Russia}}
\author{P.\,D.\,Grigoriev}  \LND \MSS
\author{A.\,M.\,Dyugaev} \LND
\author{T.\,I.\,Mogilyuk} \NRCK
\author{A.\,D.\,Grigoriev} \MSS

\begin{abstract}
It is shown that rough inner walls of a trap of ultracold neutrons can be coated with a superfluid helium film much thicker than the depth of penetration of ultracold neutrons into helium. This coating should reduce the rate of loss of ultracold neutrons caused by absorption in the walls of the trap by orders of magnitude. It is demonstrated that triangular roughness is more efficient than rectangular for the reduction of the rate of loss of ultracold neutrons. Triangular roughness is more easily implemented technically and such diffraction gratings are fabricated industrially. Other methods are proposed to increase the thickness of the protective helium film.
\end{abstract}

\maketitle

\section{Introduction}
The study of the properties of a free neutron and its interaction with known or hypothetical fields provides valuable information on elementary particles and their interactions \cite{Abele/2008,Musolf/2008,Dubbers/2011,WietfeldtColloquiumRMP2011,GONZALEZALONSO2019165}. Accurate measurements of the lifetime of the neutron $\tau_{n}$ make it possible to determine weak coupling constants, which is important for elementary particle physics, astrophysics, and cosmology \cite{Abele/2008,Musolf/2008,Dubbers/2011,WietfeldtColloquiumRMP2011,GONZALEZALONSO2019165}. Measurements of the asymmetry of the $\beta$ decay of the neutron provide information on the ratio of the vector and axial vector weak coupling constants \cite{Beam2019PhysRevLett.122.242501,PhysRevLett.105.181803,PhysRevC.101.035503}. The search for a nonzero electric dipole moment of the neutron \cite{Pospelov2005,Baker/2006,SerebrovJETPLetters2014} limits the $C P$ violation. Resonant transitions between discrete quantum energy levels of neutrons in the gravitational field of the Earth \cite{NesvizhevskyNature2002,UCNResonancePhysRevLett.112.151105} allow the study of the gravitational field at the microscale and impose constraints on dark matter. Neutron diffraction on crystals can be used to search for new internucleon interactions \cite{Voronin2018Jan,Voronin2020Nov}.

The highest accuracy of measurement of the lifetime of the neutron is currently reached with special traps for ultracold neutrons (UCNs) \cite{Serebrov2008PhysRevC.78.035505,ArzumanovPhysLettB2015,Serebrov2017,Serebrov2018PhysRevC.97.055503}. Ultracold neutrons in such traps are confined from above by the gravitational field of the Earth and from below and sides by a material that weakly absorbs neutrons and produces a potential barrier with the height $V_{0} \lesssim 300$ $\mathrm{neV}$ \cite%
{Golub/1991,Ignatovich/1990,Ignatovich1996,PhysRevLett.105.181803,PhysRevC.101.035503,Baker/2006,SerebrovJETPLetters2014,NesvizhevskyNature2002,Serebrov2008PhysRevC.78.035505,ArzumanovPhysLettB2015,Serebrov2017,Serebrov2018PhysRevC.97.055503,Review2019Pattie}. The latest measurements with such traps give \cite{Serebrov2018PhysRevC.97.055503} $\tau_{n}=(881.5$ $\pm$ $0.7$ [stat] $\pm$ $0.6$ [syst]) s. Since the neutron has a magnetic moment of $60$ $\mathrm{neV}/\mathrm{T}$, the magnetogravitational capture of UCNs is possible \cite
{Huffman2000,PhysRevC.94.045502,PhysRevC.95.035502,Ezhov2018,Pattie2018}. The most accurate measurement of the lifetime of the neutron yields $\tau_{n}=(877.7$ $\pm$ $0.7$ [stat] $+$ $0.4$ $/-$ $0.2$ [syst]) $\mathrm{s}$ \cite{Pattie2018}. According to the estimates in \cite{Serebrov2018PhysRevC.97.055503,Pattie2018}, the errors of these methods are no more than $1$ $\mathrm{s}$, but the corresponding $\tau_{n}$ values differ by almost $4$ $ \mathrm{s}$. Such a large difference is apparently due to ignored or underestimated loss of UCNs in magnetic traps. The measurement of the lifetime of the neutron with a beam of cold neutrons \cite{BeamPhysRevC.71.055502,BeamPhysRevLett.111.222501,BeamReview2020}, which is the main alternative to the measurement with UCNs, gives $\tau_{n}=(887.7$ $\pm$ $1.2$ $[\mathrm{stat}]$ $\pm$ $1.9$ $[\mathrm{syst}])$ $\mathrm{s}$. The difference of this value from the measurements by other methods is even larger, which is a known unsolved enigma. Possible reasons for this difference are actively sought from ignored errors in experiments with the neutron beams \cite{Serebrov2021PhysRevD.103.074010} to new decay channels of the neutron  \cite{BeamReview2020} or even dark matter \cite{DarkMatter2021PhysRevD.103.035014}.

The error of current experiments with traps of UCNs can also be larger than the estimate $\leq 1$ $\mathrm{s}$ because the procedure of determination of $\tau_{n}$ from experimental data involves their extrapolation by more than $15$ $\mathrm{s}$ to the limit of <<zero loss>> of neutrons in traps. One of the many possible reasons for incorrect extrapolation is the approximation of an isotropic (i.e., uniformly distributed in directions) impact of neutrons on the surface of a trap to calculate the loss rate. The real angular distribution of the velocities of UCNs is not isotropic and depends on the height: the larger the height from the bottom of the trap, the smaller the vertical component of the velocity of UCNs because of the gravitational field. This can affect both the geometric and temperature extrapolations. This effect can in principle be taken into account in Monte Carlo calculations if the normal component of the velocity of the neutron is determined for each impact with the wall. The roughness of the surface also affects the loss rate \cite{Golub/1991,Ignatovich/1990}, but it is more difficult to calculate this effect. Although an extrapolation time of about $15$ $\mathrm{s}$ is the great achievement of last 15 years \cite{Serebrov2008PhysRevC.78.035505,Serebrov2018PhysRevC.97.055503}, it is rather long and prevents a further significant increase in the accuracy of measurement of $\tau_{n}$. To reduce the extrapolation interval, it is necessary to reduce the loss rate of UCNs. The main reason for this loss is the interaction with the walls of traps, which slightly absorb neutrons.

One of the fundamental solutions to the problem of absorption of neutrons by the walls of the trap is their coating with liquid ${ }^{4} \mathrm{He}$, which does not absorb neutrons. Superfluid ${ }^{4} \mathrm{He}$ coats all the walls of a vessel with a thin film because of the van der Waals attraction. However, the thickness of this film $d_{\mathrm{He}}$ is too small. Since neutrons with energies below the potential barrier $V_{0}^{\mathrm{He}}=18.5$ $ \mathrm{neV}$ rise to the maximum height $h_{\max }=V_{0}^{\mathrm{He}} /\left(m_{n} g\right) \approx 18$ $\mathrm{cm}$, we are interested in the height above the helium level that is much larger than the capillary length $a_{\mathrm{He}}=\sqrt{\sigma_{\mathrm{He}} /\left(g \rho_{\mathrm{He}}\right)}=0.5$ $\mathrm{mm}$, where $\sigma_{\mathrm{He}}=0.354$ $\mathrm{dyn} / \mathrm{cm}$ is the surface tension of ${ }^{4} \mathrm{He}$, $g=9.8$ $\mathrm{m}/ \mathrm{s}^{2}$, and $\rho_{\mathrm{He}} \approx 0.145$ $\mathrm{g}/\mathrm{cm}^{3}$ is the density of liquid ${ }^{4} \mathrm{He}$. The film thickness at this height is $d_{\mathrm{He}} \approx 10$ $\mathrm{nm}$, which is much smaller than the penetration depth $\kappa_{0 \mathrm{He}}^{-1}=\hbar / \sqrt{2 m_{n} V_{0}^{\mathrm{He}}} \approx 33.5$ $\mathrm{nm}$ of neutrons into ${ }^{4} \mathrm{He}$. Hence, such a film hardly protects neutrons from absorption inside the wall of the trap.

The problem of an increase in the thickness of the helium film is important only for the vertical (side) walls of traps of UCNs. The bottom of the trap can be easily coated with the helium film with the required thickness $d_{\mathrm{He}} \gg \mathrm{\kappa}_{0 \mathrm{He}}^{-1} \approx 33.5$ $\mathrm{nm}$; neutrons are confined from above by the gravitational field of the Earth. Only the lower part of the side walls to the height $h<a_{\mathrm{He}} \sqrt{2} \ll h_{\max }$ are coated with the meniscus with the thickness $d_{\mathrm{He}} \sim a_{\mathrm{He}} \geqslant \kappa_{0 \mathrm{He}}^{-1}$. To increase the thickness of the helium film on the side walls of the trap above the capillary length, it was proposed to store neutrons in a rotating vessel with helium \cite{Bokun/1984,Alfimenkov/2009}. However, the reflection of neutrons from the moving surface gradually increases their kinetic energy, which finally exceeds the potential barrier $V_{0}^{\mathrm{He}}=18.5$ $\mathrm{neV}$, and a neutron leaves the trap. Furthermore, the rotating liquid generates additional bulk and surface excitations, which increases the inelastic scattering rate of neutrons. Consequently, time-independent coating of the walls of the trap with liquid ${ }^{4} \mathrm{He}$ is necessary.

We recently assumed \cite{Grigoriev2021Nov} that the thickness of the helium film on the rough wall of the trap of UCNs
increases because of capillary effects. The roughness of the wall usually increases the loss of neutrons by a factor of $2-3$ through absorption inside the walls of the trap because this roughness makes the average repulsive potential of the walls smoother, so that the wavefunction of the neutron penetrates deeper into the wall  \cite{Golub/1991,Ignatovich/1990}. However, the average thickness of the helium film deposited on such rough wall can be strongly increased compared to the smooth surface because of capillary effects, thus reducing the loss of neutrons. Indeed, the roughness of the wall increases its area and, correspondingly, the role of capillary effects. If the scale of the surface roughness is $l_{\mathrm{R}} \ll a_{\mathrm{He}}$,
to minimize the surface tension energy, the helium film even on the rough wall should have an almost flat interface with vacuum. Therefore, superfluid helium fills all small cavities with the size $l_{\mathrm{R}} \lesssim a_{\mathrm{He}}$ in the wall.
In this work, we develop this idea and propose implementable variants of the side walls of the trap, for which the thickness of the helium film is large and the loss rate of UCNs through absorption in the wall of the trap decreases by orders of magnitude.

\section{Helium film on the rough surface: general formulas}

To describe the profile of the helium film on the rough surface, it is necessary to minimize the energy functional of this film
\begin{gather}
E_{\mathrm{tot}}=V_{\mathrm{g}}+E_{\mathrm{s}}+V_{\mathrm{w}}.  \label{eq:EtotM}
\end{gather}
Here, $V_{g}$ is the gravity term given by the expression
\begin{gather}
V_{\mathrm{g}}=\rho_{\mathrm{He}} g \int z d_{\mathrm{He}}\left(\mathbf{r}_{\|}\right) d^{2} \mathbf{r}_{\|}, \label{eq:Vg}
\end{gather}
where $\mathbf{r}_{\|}=\{x,\,z\}$ is the two-dimensional vector of the horizontal, $x$, and vertical, $z$, coordinates on the wall,
\begin{gather}
d_{\mathrm{He}}\left(\mathbf{r}_{\|}\right)=\xi\left(\mathbf{r}_{\|}\right)-\xi_{\mathrm{W}}\left(\mathbf{r}_{\mathrm{W}}\right) \label{eq:dHeDef}
\end{gather}
is the thickness of the helium film depending on the coordinates, and $\xi(\mathbf{r_{||}})$ and $\xi_{\mathrm{W}}\left(\mathbf{r}_{\|}\right)$ are the functions that describe the profiles of the He surface and the wall of the trap. The gravity term always reduces the thickness of the helium film. Below, we consider the roughness of the wall with the characteristic length scale $a_{\mathrm{He}} \ll h_{\max }$. The variation of the $z$ coordinate at this small length scale can be neglected compared to its average value $\langle z\rangle$ equal to the height $h$ of the roughness above the helium level in the trap. Therefore, the $z$ coordinate in Eq. (\ref{eq:Vg}) can be replaced by the height $h$.

The second term $E_{\mathrm{s}}$ in Eq. (\ref{eq:EtotM}) describes the surface tension energy and is given by the formula
\begin{gather}
E_{\mathrm{s}}=\sigma_{\mathrm{He}} \int \sqrt{1+\left[\boldsymbol{\nabla} \xi\left(\mathbf{r}_{\|}\right)\right]^{2}} d^{2} \mathbf{r}_{\|}.  \label{eq:Es}
\end{gather}
Its root dependence complicates the problem of determination of the exact surface profile $\xi(\mathbf{r_{||}})$. As a rule, this problem can be solved analytically only in the limit of
small curvature of the surface $|\nabla \xi(\mathbf{r}_{||} )| \ll 1$, when the approximation $\sqrt{1+\left[\boldsymbol{\nabla} \xi\left(\mathbf{r}_{\|}\right)\right]^{2}} \approx 1+[\boldsymbol{\nabla} \xi(\mathbf{r_{||}})]^{2} / 2$ is valid. In our case, the condition $|\nabla \xi(\mathbf{r_{||}})| \ll 1$ is not necessarily fulfilled and we do not use this approximation for qualitative estimates. We instead search for the minimum of the initial functional specified by Eqs. (\ref{eq:EtotM})--(\ref{eq:Es}) in the class of trial functions.

The gravity and surface tension terms in Eq. (\ref{eq:EtotM}) are important at the macroscopic length scale $a_{\mathrm{He}}$. The van der Waals term $V_{W}$ describing the attraction of helium to the material of the wall is noticeable only at
much smaller distances $d_{\mathrm{He}}^{\mathrm{min}} \approx 10$ nm $\ll a_{\mathrm{He}}$. Since the range of van der Waals forces $d_{\mathrm{He}}^{\min}$ is five orders of magnitude smaller than the capillary length $a_{\mathrm{He}}$, the effect of the gravitational force and surface tension of the helium surface on $V_{W}$ can be neglected and theoretical analysis is simplified. The van der Waals potential $V_{\mathrm{W}}$ for the smooth wall depends only on the material of the wall and the film thickness: $V_{\mathrm{W}}=V_{\mathrm{W}}\left(d_{\mathrm{He}}\right)$. In the absence of surface tension, van der Waals forces would result in the coating of the rough surface with the helium film of the thickness $d_{\mathrm{He}} \sim d_{\mathrm{He}}^{\min }$, which almost repeats the wall profile if the roughness scale is $l_{\mathrm{R}} \gg d_{\mathrm{He}}^{\min }$. Thus, only the first two terms can be retained in the functional $E_{\mathrm{tot}}[\xi(\mathbf{r_{||}})]$ given by Eq. (\ref{eq:EtotM}), and the effect of the van der Waals term $V_{\mathrm{W}}$ is reduced to the <<boundary conditions>> of the minimum thickness of the helium film $d_{\mathrm{He}} \geq d_{\mathrm{He}}^{\min } \approx 10$ $\mathrm{nm}$.

Such a minimum helium film with the thickness $d_{\mathrm{He}} \sim d_{\mathrm{He}}^{\mathrm{min}}$ caused by the van der Waals attraction provides an additional surface tension energy $\Delta E_{\mathrm{s}}$, which can be even larger than the addition $\Delta V_{\mathrm{g}}$ to the gravity term appearing because of an additional helium amount necessary for the helium surface to be flat: $\xi(\mathbf{r_{||}})=$ const $=\max \left\{\xi_{W}(\mathbf{r_{||}})\right\}+d_{\mathrm{He}}^{\min }$. This additional helium amount depends on the wall roughness profile and can strongly increase the average thickness of helium films. This increase in the effective thickness of the helium film caused by capillary effects can apparently explain the difference in its experimental values determined by different methods \cite{Atkins1950,BurgeJackson1951,RaymondBowers,Atkins1957,PhysRevA.7.790,PhysRevA.9.1312}. Indeed, the total weight of helium is measured by the microweighing method \cite{RaymondBowers}. This amount of helium includes filled cavities on the surface, which were not detected by optical methods in \cite{BurgeJackson1951,Ham1954}. Consequently, the microweighing method gives a thicker helium film on the rough surface \cite{RaymondBowers}. Since the typical thickness of the helium film caused by the van der Waals forces is small, $d_{\mathrm{He}} \sim 10$ $ \mathrm{nm}$, even miniature surface roughness with a height of about $10$ $\mathrm{nm}$ can strongly affect the measured values of the thickness of the helium film.

\section{Helium film on the wall with modulation in the form of a triangular wave}


\begin{figure}[!htb]
	\includegraphics[width=0.48\textwidth]{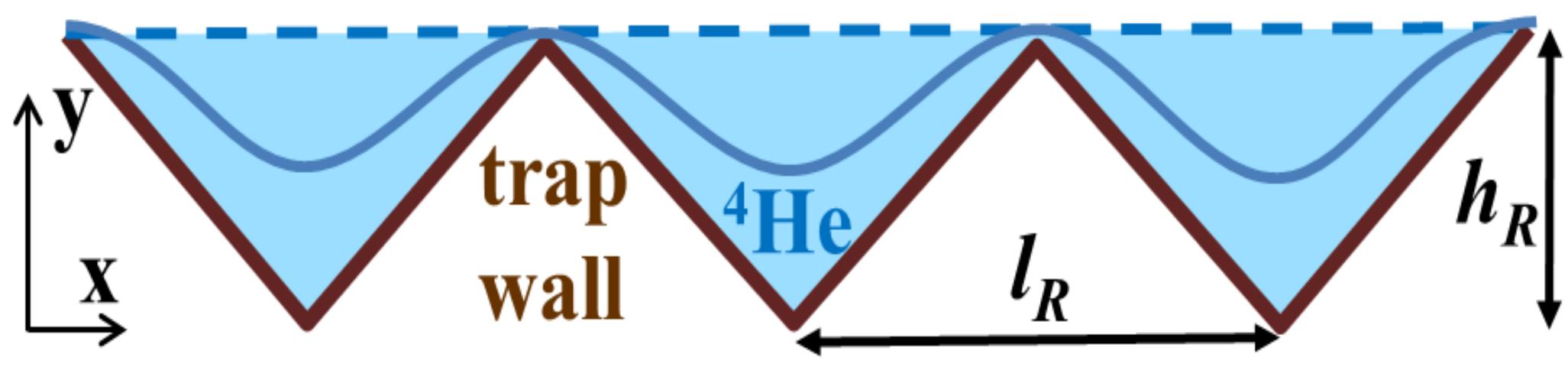}
	\caption{Color online) Wall with (brown polyline) triangular roughness coated with liquid helium with (blue dashed line) almost flat and (blue solid line) periodically modulated surfaces.}
	\label{fig:FigTriang}
\end{figure}

We consider the helium film on the rough wall in the form of a one-dimensional triangular wave with the period $l_{\mathrm{R}}$ and depth $h_{\mathrm{R}}$, as shown in Fig. \ref{fig:FigTriang}. To approximately estimate the necessary roughness parameters of such a wall, we compare the energies of a very thin film with a thickness of $\sim d_{\mathrm{He}}^{\min }$ repeating the surface relief and the helium film with the flat surface shown by the blue dashed line in Fig. \ref{fig:FigTriang}. Since the thickness of the film is larger than $d_{\mathrm{He}}^{\mathrm{min}}$, these two configurations have approximately the same van der Waals energy. Their gravitational energies (\ref{eq:Vg}) per unit area of the wall is $\Delta V_{\mathrm{g}}=-g\rho_{\mathrm{H}}h_{\mathrm{R}}h/2$. The difference of surface tension energies (\ref{eq:Es}) of these two configurations is equal to the product of the difference of their surface areas and the surface tension of liquid helium $\sigma_{\mathrm{He}}$; this difference per unit surface area is $\Delta E_{\mathrm{s}}=\sigma_{\mathrm{He}}\left(\sqrt{1+\left(2 h_{\mathrm{R}} / l_{\mathrm{R}}\right)^{2}}-1\right) .$ The sum $\Delta V_{\mathrm{g}}+\Delta E_{\mathrm{s}}$ is positive; i.e., the flat free surface of the helium film is more favorable than that repeating the wall relief if two conditions are satisfied: (i) the modulation period is bounded from above:
\begin{gather}
l_{\mathrm{R}}<l_{R}^{\max }=4 \sigma_{\mathrm{He}} /\left(g h \rho_{\mathrm{He}}\right)=4 a_{\mathrm{He}}^{2} / h, \label{eq:lR}
\end{gather}
and (ii) the modulation depth is bounded from below:
\begin{gather}
h_{\mathrm{R}}>h_{\mathrm{R}}^{\min }=\frac{4 a_{\mathrm{He}}^{2} / h}{\left(4 a_{\mathrm{He}}^{2} /\left(l_{\mathrm{R}} h\right)\right)^{2}-1}=\frac{l_{\mathrm{R}} \eta}{1-\eta^{2}}, 	\label{eq:hR}
\end{gather}
where
\begin{gather}
\eta \equiv l_{\mathrm{R}} h /\left(4 a_{\mathrm{He}}^{2}\right). 	\label{eq:eta}
\end{gather}
According to Eq. (\ref{eq:lR}), $\eta<1$.
The substitution of the value $a_{\mathrm{He}}=0.5$ $ \mathrm{~mm}$ and the maximum height $h_{\max} \approx 18$ $\mathrm{cm}$ to which neutrons with energies lower than $V_{0}^{\text{He}}$ can rise into Eq. (\ref{eq:lR}) yields the maximum roughness period $l_{\mathrm{R}}^{\max }\left(h_{\max }\right) \approx 5.6$ $\mu \mathrm{m}$. Since $l_{\mathrm{R}}^{\max } \propto 1 / h$, the modulation period at a smaller height can be larger. A part of the kinetic energy of the neutron that is determined only by the vertical component of its velocity $v_{z}$ is on average one-third of its total energy and one-fifth of the maximum energy $V_{0}^{\mathrm{He}}$. Consequently, the neutron usually collides with the vertical wall at a height below $h_{\operatorname{mad}} / 5$, where the roughness period can be made larger by a factor of 5 : $l_{R}<l_{R}^{\max }\left(h_{\max}/ 5\right) \approx 28$ $\mu \mathrm{m}=820 \kappa_{0 H e}^{-1}.$

According to $\mathrm{Eq}$. (\ref{eq:lR}), the minimum roughness amplitude of the wall at $\eta \propto 1$ is $h_{\mathrm{k}}^{\min } \propto l_{\mathrm{R}}^{2}$. Therefore, at a smaller modulation period $I_{\mathrm{R}}$, a less sharp roughness can be taken and, correspondingly, its brittleness is smaller. If the value $l_{\mathrm{R}}=l_{\mathrm{R}}^{\max}/2$ is taken for reliability (in the case of strong fluctuations of $l_{\mathrm{R}}$), which certainly satisfies $\mathrm{Eq}$. (\ref{eq:lR}), $\mathrm{Eq}$. (\ref{eq:hR}) takes the form $h_{\mathrm{R}}>2 l_{\mathrm{R}} / 3 ;$ i.e., modulation should be quite deep. If $l_{\mathrm{R}}=l_{\mathrm{R}}^{\max } / 4=a_{\mathrm{He}}^{2} / h$, which satisfies $\mathrm{Eq}$. (\ref{eq:lR}) with even a larger margin, Eq. (\ref{eq:hR}) has the form $h_{\mathrm{R}}>4l_{\mathrm{R}}/15;$ i.e., a smaller modulation depth can be taken and the roughness will be less brittle.

\begin{figure}[!tbh]
	\includegraphics[width=0.48\textwidth]{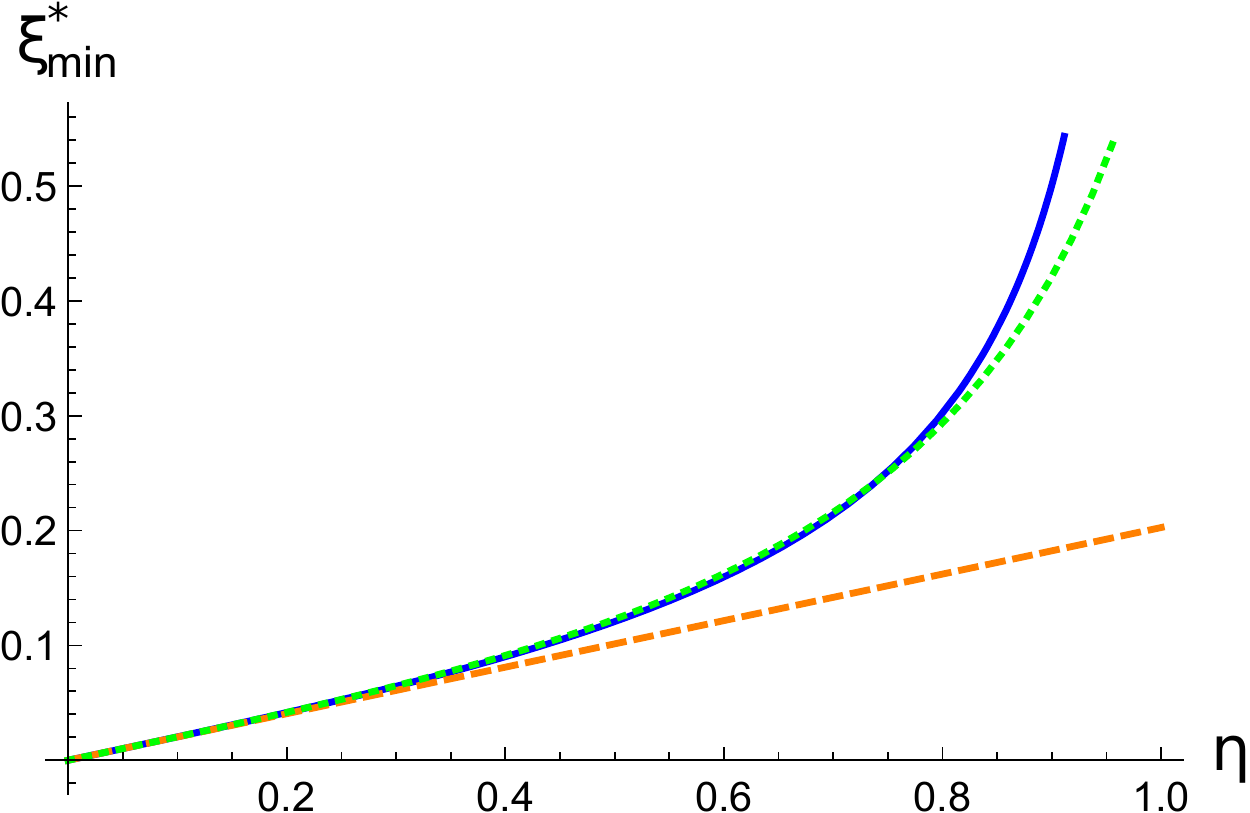}
	\caption{(Color online) Parameter $\xi_{\min }^{*} \equiv \xi_{0} / l_{\mathrm{R}}$ versus $\eta=l_{\mathrm{R}} h /\left(4 a_{\mathrm{He}}^{2}\right)$ obtained (blue solid line) by minimizing Eq. (\ref{eq:eta}), (orange dashed line) in the approximation $\left|\xi^{\prime}(x)\right| \ll 1$, and (green dashed line) in the analytical approximation from Eq. (\ref{eq:xiApp}).}
	\label{fig:FigXi}
\end{figure}

The two profiles of the helium film considered above are certainly not optimal, i.e., not corresponding to the minimum of the total energy (\ref{eq:EtotM}). When the curvature of the surface is small, $|\nabla \xi(\mathbf{r}_{||})| \ll 1$, the gravity term (\ref{eq:Vg}) is linear in $\xi$, whereas the capillary term (\ref{eq:Es}) is quadratic in $\xi$. Consequently, the gravity term prevails at small $\xi$ values and, therefore, at least small curvature of the surface always exists and the flat surface $\xi(\mathbf{r}_{||})=$ const is impossible. At the same time, the repetition of the triangular wall profile by the free helium surface is also not profitable because it has corners. The real surface of this relief wall is described by a smooth periodic function $\xi(x)$ with the period $l_{\mathrm{R}}$, which is schematically shown by the blue solid line in Fig. \ref{fig:FigTriang}. Therefore, it is possible to take the approximate trial function
\begin{gather}
\xi(x)=\xi_{0} \cos \left(2 \pi x / l_{\mathrm{R}}\right) 	\label{eq:xi}
\end{gather}
and to determine the amplitude $\xi_{0}$ at which the total energy (\ref{eq:EtotM}) has a minimum. From Eqs. (\ref{eq:xi}) and (\ref{eq:Vg}), the gain in the gravitational energy because of such sinusoidal curvature of the surface per unit area of the wall is determined in the form
\begin{gather}
	\Delta V_{\mathrm{g}}=\frac{\rho_{\mathrm{He}} g h \xi_{0}}{l_{\mathrm{R}}} \int_{-l_{\mathrm{R} / 2}}^{l_{\mathrm{R} / 2}}\left[\cos \left(2 \pi x / l_{\mathrm{R}}\right)-1\right] d x \label{eq:DVg} \\\nonumber
	=-\rho_{\mathrm{He}} g h \xi_{0}. 
\end{gather}
The substitution of Eq. (\ref{eq:xi}) into Eq. (\ref{eq:Es}) gives the following expression for curvature-induced loss in the surface tension energy per unit area of the wall:
\begin{gather}
	\Delta E_{\mathrm{s}}=\frac{\sigma_{\mathrm{He}}}{l_{R}} \int_{-l_{\mathrm{R}}/2}^{l_{\mathrm{R}}/ 2}\left(\sqrt{1+\left[\xi^{\prime}(x)\right]^{2}}-1\right) d x \\
	=\sigma_{\mathrm{He}}\left(2 E\left[-2 \pi i \xi_{0} / l_{\mathrm{R}}\right] / \pi-1\right), 	\label{eq:DEs}
\end{gather}
where $E[x]$ is the complete elliptic integral of the second kind. The sum of Eqs. (\ref{eq:DVg}) and (\ref{eq:DEs}) gives the change in the total energy caused by sinusoidal curvature of the surface
\begin{gather}
\Delta E_{\mathrm{tot}}=\rho_{\mathrm{He}} g a_{\mathrm{He}}^{2}\left(2 E\left[2 \pi i \xi^{*}\right] / \pi-1-4 \eta \xi^{*}\right), 	\label{eq:DEtot}
\end{gather}
where $\xi^{*} \equiv \xi_{0} / l_{\mathrm{R}}$ is the normalized wave amplitude, which is determined by minimizing the total energy (\ref{eq:eta}). It is seen that the position of the minimum $\xi_{\mathrm{min}}^{*}$ depends only on the parameter $\eta=l_{\mathrm{R}} h /\left(4 a_{\mathrm{He}}^{2}\right)$ given by Eq. (\ref{eq:DEtot}). This parameter continuously appears in our problem. The dependence $\xi_{\min }^{*}(\eta)$ obtained by minimizing Eq. (\ref{eq:eta}) is shown by the blue solid line in Fig. 2. It diverges at $\eta \rightarrow 1$. Thus, we again arrive at Eq. (\ref{eq:lR}) determining the maximum modulation period of the wall roughness. The function $\xi_{\min }^{*}(\eta)$ at $\eta<0.8$ is satisfactorily approximated by the formula (see Fig. \ref{fig:FigXi})
\begin{gather}
\xi_{\min }^{*}(\eta) \approx\left(2 / \pi^{2}\right) \eta /\left(1-0.7 \eta^{2}\right) 	\label{eq:xiApp}
\end{gather}
The wave $\xi(x)$ specified by Eq. \ref{fig:FigXi} does not touch the solid wall with the triangular profile, which is shown in Fig. 1, at
\begin{gather}
h_{\mathrm{R}}>h_{\mathrm{R}}^{\min } \approx 2.3 \xi_{0}^{\min } \equiv 2.3 l_{\mathrm{R}} \xi_{\min }^{*}(\eta). 	\label{eq:hRs}
\end{gather}
Formula (\ref{eq:hRs}) and Fig. \ref{fig:FigXi} determine the minimum depth of the triangular relief of the wall at which it does not touch the free helium surface anywhere except for bulges at integer values of $x / l_{\mathrm{R}}$. This condition on $h_{\mathrm{R}}$ is weaker by approximately a factor of 2 than the condition (\ref{eq:hR}) because the curved surface of the helium film described by Eq. (\ref{eq:xi}) with a small amplitude $\xi_{0}$ insignificant for the film to touch the solid wall is more favorable than the absolute flat surface. This weakening of the condition (\ref{eq:hR}) on the roughness depth $h_R$
facilitates the practical implementation of such a wall. Nevertheless, the condition (\ref{eq:lR}) on the roughness period does not change.

If the condition $|\nabla \xi(\mathbf{r}_\|)| \ll 1$ corresponding to $\xi_{0} \ll$ $l_{\mathrm{R}}$ were used from the beginning, we would obtain the linear function $\xi_{\min }^{*}(\eta)=2 \eta / \pi^{2}$ shown by the orange dashed line in Fig. 2. According to Fig. 2, this linear approximation is applicable at $\eta<0.4$, and it is completely inapplicable at $\eta \rightarrow 1$.

The triangular roughness of the wall shown in Fig. 1 has some significant advantages compared to the rectangular profile shown and considered in \cite{Grigoriev2021helium} (Fig. 6).

First, the triangular wall with the required parameters will be much less brittle than the rectangular one. It touches the free helium surface only at bulges at $x=$ $x_{n}=l_{\text{R}} n$, where $n$ is an integer. The thickness of the helium film at these points decreases to $d_{\mathrm{He}}^{\min } \sim 10 \mathrm{~nm}$. However, since the derivative at these points is zero, $\left.\xi^{\prime}(x)\right|_{x=x_{n}}=0$, the thickness of the film increases linearly with the distance $\Delta x=x-x_{n}$ from these points because of the linear dependence $\zeta_{\text{W}}(\Delta x)$ :
\begin{gather}
d_{\mathrm{He}}(x) \approx d_{\mathrm{He}}^{\min }+\left|x-x_{n}\right|\left(2 h_{\mathrm{R}} / l_{\mathrm{R}}\right). 	\label{eq:dHexn}
\end{gather}
According to the approximate semiclassical formula [37] of the exponential decrease in the wavefunction of the neutron inside the helium film, the helium film reduces the absorption of the neutrons by the material of the wall at the point $x$ by a factor of
\begin{gather}
\gamma(x)=\left[\psi(0) / \psi\left(d_{\mathrm{He}}(x)\right)\right]^{2} \sim \exp \left[-2 \kappa_{\mathrm{He}} d_{\mathrm{He}}(x)\right],  \label{eq:psiS1}
\end{gather}
where $\kappa_{\mathrm{He}} \approx \hbar / \sqrt{2 m_{n}\left(V_{0}^{\mathrm{He}}-E_{\mathrm{kin} \perp}\right)}$. Since the neutron velocity component normal to the wall is on average responsible for only one-third of its total kinetic energy $\bar{E}_{\mathrm{kin} \perp} \approx \bar{E}_{\mathrm{kin}} / 3 \approx V_{0}^{\mathrm{He}} / 5$, the average value is $\kappa_{\mathrm{He}} \approx h / \sqrt{2 m_{n}\left(V_{0}^{\mathrm{He}}-\bar{E}_{\mathrm{kin} \perp}\right)} \approx 0.9 \kappa_{0 \mathrm{He}}$. To estimate the average $\gamma$ value, we integrate Eq. (\ref{eq:psiS1}) over the period $l_{\mathrm{R}}$ with the function (\ref{eq:dHexn}):
\begin{gather}
	\bar{\gamma} \approx \frac{2}{l_{\mathrm{R}}} \int_{0}^{l_{\mathrm{R}} / 2} d x \exp \left[-2 \kappa_{\mathrm{He}}\left(d_{\mathrm{He}}^{\min }+2 x h_{\mathrm{R}} / l_{\mathrm{R}}\right)\right] \\
	=\frac{\exp \left(-2 \kappa_{\mathrm{He}} d_{\mathrm{He}}^{\min }\right)}{2 \kappa_{\mathrm{He}} h_{\mathrm{R}}} \approx \frac{0.3}{\kappa_{\mathrm{He}} h_{\mathrm{R}}} \ll 1 . 	\label{eq:gammaAv}
\end{gather}
According to this estimate, roughness with the characteristic depth $h_{\mathrm{R}}>30 \kappa_{0 \text{He}}^{-1} \approx 1$ $\mu \mathrm{m}$ and period $l_{\mathrm{R}}<$ $l_{\mathrm{R}}^{\max }(h) \approx\left(h_{\max } / h\right) \times 5.6$ $\mu \mathrm{m}$ can reduce the absorption of neutrons by the wall by a factor of 100. Thus, $h_{\mathrm{R}} \ll$ $l_{\mathrm{R}}$ and such roughness will be stable against mechanical load. This property is an advantage of the proposed triangular surface relief compared to the rectangular relief considered in \cite{Grigoriev2021Nov}, where to reduce the absorption of neutrons by the wall, roughness satisfying the relation $h_{\mathrm{R}}>l_{\mathrm{R}}$, i.e., a very brittle wall relief, was required. It is noteworthy that diffraction gratings with the period $l_{\mathrm{R}} \approx 4$ $\mu \mathrm{m}$ and depth $h_{\mathrm{R}} \approx 0.2$ $\mu \mathrm{m}$ are already actively used for the scattering of UCNs \cite{Kulin2016Mar,Kulin2019Nov}.

Second, the proposed triangular profile corresponds to standard diffraction gratings, whose production technology was developed long ago and is used industrially. Consequently, such an artificial roughness can be easily created \footnote{Diffraction gratings with the period $l_{\mathrm{R}}=1$ $\mu$ $\mathrm{m}$, sizes of $1.524$ $\times$ $0.1524$ $\mathrm{m}$, and required parameters are available at a price of $\$ 20$ at www.amazon.com. Gratings with the triangular profile and a much smaller period $l_{R} \geq 150$ $\mathrm{nm}$ are available, eg., at nilt.com.}. Electron beam lithography can provide a much smaller period of the diffraction grating $l_{R} \leq 100$ nm \cite{Bourgin2010}. Thus, the triangular surface roughness profile makes our idea of the coating of walls of traps of UCNs with a sufficiently thick film of superfluid ${ }^{4} \mathrm{He}$ practically implementable.
\section{Other methods}
Well-developed \textit{powder metallurgy} methods can be used to obtain the required rough/porous surface of the trap for the purpose of its further coating with the helium film. The simple pressing of a pure fine-grained beryllium powder at pressures of about $1.0$--$1.5$ GPa and temperatures of about $1000^{\circ}$ C will already give the surface profile applicable to hold the required amount of liquid helium on the vertical wall of the trap.

Moreover, efficient methods of obtaining pure porous beryllium were developed already in 1998 in \cite{BePorousPatentEn} to coat the walls of nuclear fusion reactors. These methods can provide a large amount of beryllium with a controlled uniform porosity ranging from 10 to $70 \%$ and $100 \%$ open porosity \cite{BePorousPatentEn}.

The rough surface of traps of UCNs can also be obtained by one of the standard methods of sintering of small particles (dust) of a material weakly absorbing neutrons. In this case, an important requirement is a low concentration of impurity particles.

The coating of the walls of the trap of UCNs with a diamond nanopowder often used in experiments with UCNs \cite{DiamondPowder} can also provide the required roughness of the surface to make the helium film sufficiently thick.
The superfluid ${ }^{4} \mathrm{He}$ film protecting neutrons from absorption in the walls of the trap can be obtained using a \textit{thin metallic wire} wound (entangled) to create a sufficiently large surface area $S$ per unit volume $V$ but with a small volume fraction $\phi_{\text {wire}}=V_{\text {wire}} / V$. Such a configuration of entangled wires can be suspended from above to hang loose under the action of the gravitational force. Such a wire should be made of a material weakly absorbing neutrons, e.g., beryllium. In this case, an insignificant fraction of the wire $\phi_{wire}\sim0.1$ completely coated with ${}^4$He does not strongly absorb
neutrons.

The surface and volume of the wire with the diameter $d$ and length $L$ per unit volume $V$ are $S_{\text {wire }}=L \pi d$ and $V_{\text {wire }}=L \pi d^{2} / 4$, respectively. Owing to capillary effects, superfluid ${ }^{4} \mathrm{He}$ will rise through such randomly entangled wires to the height $h$ at which the gravitational and surface energies of liquid helium are equal to each other: $h g V \rho_{\mathrm{He}}\approx S \sigma_{\mathrm{He}}$. This imposes a constraint on the length $L$ and volume fraction $\phi_{\text {wire }}$ of the wire:
\begin{gather}
S=L \pi d>h g V \rho_{\mathrm{He}} / \sigma_{\mathrm{He}}=h V / a_{\mathrm{He}}^{2}, \label{eq:Sw}
\end{gather}
which corresponds to the relation
\begin{gather}
\phi_{\text {wire}}>h d /\left(4 a_{\text {He }}^{2}\right) \text {. } \label{eq:phiw}
\end{gather}
The substitution of $h=h_{\max }=18$ $\mathrm{cm}$ into $\mathrm{Eq}$. (\ref{eq:phiw}) gives an upper bound for the thickness of the wire
\begin{gather}
d<d_{\max }=4\phi_{\text{wire}}a_{\mathrm{He}}^{2} / h_{\max } . \label{eq:dw}
\end{gather}
The condition $\phi_{\text{wire}}\leq 0.1$ means that the wire should be thinner than $d_{\max }=0.56$ $\mu \mathrm{m}$. Under a more stringent condition $\phi_{\text{wire}}\leq 0.01,\, d_{\max}=14$ nm. It is likely more difficult technically to fabricate such a thin wire with a sufficient total length than to use the triangular roughness of the surface (diffraction grating) considered above. Entangled segments of the wire with a much smaller thickness of about 5 $\mathrm{nm}$ can be fabricated by the aggregation of nanoparticles in ${ }^{4}\mathrm{He}$ vortices \cite{Gordon_2015,Karabulin2021Feb}, but it is difficult to control the length and concentration of such wire segments by this method.
\section{Conclusions}
The main channel of loss of UCNs in material traps is the absorption of neutrons by the walls of traps. Neutrons are not absorbed by ${ }^{4} \mathrm{He}$. Therefore, the storage time of UCNs in a trap whose walls are coated with a helium film can be increased by several orders of magnitude because the inelastic scattering rate of UCNs on surface and bulk excitations of liquid helium is sufficiently low at $T<0.4$ K \cite{Grigoriev2016Aug}. However, because of van der Waals forces, the thickness of the superfluid helium film coating the vertical surfaces is $d_{\mathrm{He}}=$ $10$ nm, which is much smaller than the penetration depth $\mathrm{\kappa}_{0 \mathrm{He}}^{-1}=\hbar / \sqrt{2 m_{n} V_{0}^{\mathrm{He}}}\approx33.5$ nm of neutrons in ${ }^{4} \mathrm{He}$. Consequently, such a film does not protect neutrons from absorption inside the wall of the trap. In this work, we have proposed a technically simple method of increasing the thickness of the helium film coating the inner surface of the trap of UCNs. It is necessary to make the walls of the trap rough, e.g., in the form of a standard diffraction grating with a triangular profile. Estimates of the necessary parameters of such roughness indicate that such a rough surface can be technically obtained by various inexpensive methods.
\bibliographystyle{unsrt}
\bibliography{NeutronReviews,UCN,HeFilm}
\end{document}